\documentstyle[aps,twocolumn,eqsecnum,epsf]{revtex}
\begin{document}
\title{
\rightline{\rm OU-TAP 114}
\rightline{\rm UAB-FT 484}
The use of new coordinates for the template space in hierarchical search for 
gravitational waves from inspiraling binaries 
} 
\author{\large
 Takahiro Tanaka$^{a,b,*}$
 and 
 Hideyuki Tagoshi$^{a,\dag}$
} 

\address{\vspace{5mm}
${}^a$ Department of Earth and Space Science, Graduate 
School of Science\\
Osaka University, Toyonaka 560-0043, Japan} 
\address{${}^b$ IFAE, Departament de F{\'\i}sica, 
  Universitat Aut{\`o}noma de Barcelona\\
08193 Bellaterra $($Barcelona$)$, Spain}

\address{$^{*}$\sf tama@vega.ess.sci.osaka-u.ac.jp\\ 
$^{\dag}$tagoshi@vega.ess.sci.osaka-u.ac.jp}

\address{\vspace*{5mm}
\centerline{\rm 
\begin{minipage}{16cm}\hspace*{5mm}
We discuss a method to analyze 
data from interferometric gravitational wave detectors 
focusing on the technique of hierarchical search 
to detect gravitational waves from inspiraling binaries. 
For this purpose, we propose new coordinates 
to parameterize the template space. 
Using our new coordinates,
we develop several new techniques for two step search,  
which would reduce the computation 
cost by large amount. 
These techniques become more important
when we need to implement a $\chi^2$-test as a 
detection criterion.
\end{minipage}}
}

\maketitle
\thispagestyle{empty}
\section{Introduction}
The interferometric gravitational wave detectors, such as, 
LIGO, VERGO, GEO600 and TAMA300\cite{tamaconf}, 
are now under construction. 
Especially, TAMA300 has already done 
the first large scale data acquisition for three days 
in September 1999 \cite{tamaconf}. 
Primary targets of these detectors are inspiraling 
binary neutron stars or black holes. 
These compact binaries can be produced as a consequence of 
normal stellar evolution in binaries. 
It is also suggested that 
they might also have been produced in the early universe. 
The analysis of the first 2.1yr of photometry of 8.5 
million stars in the Large Magellanic Cloud by the MACHO 
collaboration suggests that $0.62^{+0.3}_{-0.2}$ of 
halo consists of MACHOs of mass $0.5^{+0.3}_{-0.2}M_{\odot}$ 
in the standard spherical flat rotation halo model\cite{macho}. 
If these MACHOs are black holes, it is reasonable to 
consider that they were produced in the early universe, 
and some of them are in binaries which 
coalesce due to the gravitational radiation reaction\cite{bhmacho}. 
%If an event occurs in the halos of the near-by galaxies, 
%their coalescence will produce strong gravitational waves. 
%They would be detected by above detectors even at the initial stage 
%at which the sensitivity is still low. 
Thus, it is expected that 
the observation of gravitational waves gives a 
definite answer to the question whether these MACHOs 
consist of primordial black holes or not.  

To search for 
gravitational waves 
emitted by these binaries, 
the technique of matched filtering 
is considered to be the best method.
In this method, 
detection of signals and estimation of binary's
parameters 
are done by taking the cross-correlation between 
observed data and predicted wave forms. 
For this purpose, we need to prepare theoretically predicted 
wave forms, often called {\it templates}. 
Generally, such templates depend on binary's
parameters such as 
mass, spin, coalescing time, phase, and so on. 
Since these parameters are continuous, 
what we really have to do is to 
prepare a template bank which consists of a finite number 
of representative templates. 

In this paper, 
we propose a new parameterization of 
two mass parameters of binaries. 
We show that the use of them has various advantages in performing the
matched filtering.
These parameters define two dimensional coordinates  
on the parameter space of templates. 
We can introduce a distance between two templates 
by using the cross-correlation between them. 
This distance defines a metric on the template space \cite{owen}. 
We shall show that the metric in terms of our new parameters  
approximately becomes a flat Euclidean metric. 
Thus, it becomes very simple in these coordinates 
to determine the grid points corresponding to the 
bank of templates. 
The method how to construct the new coordinates is 
explained in the succeeding section. 

Requiring that the grid in the template space is  
sufficiently fine so as not to lose real events,  
the number of templates to be searched 
tends to be very large. Especially, if we 
lower the minimum mass of binaries to be searched
$m_{min}$, the number of templates increases as 
$m_{min}^{-8/3}$\cite{owen}. 
When we search for 
gravitational waves from binary black hole 
MACHOs, we need to choose $m_{min}$  
sufficiently lower than the predicted mass of MACHOs $\sim 0.5M_{\odot}$. 
For example, let us consider that the grid is chosen so that the 
correlation between nearest neighboring templates 
becomes $0.97$. 
Then, in order to search for binaries composed of 
compact stars in the mass range between $0.2 M_{\odot}$ 
and $10M_{\odot}$, the necessary number of templates 
becomes $\sim 2\times 10^5$ for the ``TAMA noise curve''\cite{owen}. 
The matched filtering with 
the sampling rate of 3000Hz requires 
the data processing speed faster than 
80G FLOPS (FLoating Operations Per Second) for the on-line analysis. 
Now, such a powerful computation environment may be 
available. However it is still very expensive. 
Furthermore, 
there are various factors in real data analysis 
which increase the computation cost. 
One of them is the non-Gaussian nature of the detector noise,  
which we shall discuss in this paper. 
Hence, the computation cost can be much larger 
than that estimated in an ideal situation.
Thus, it is required to develop 
some methods to reduce the computation cost. 

The technique of hierarchical search is thought to be 
a promising way to realize such reduction 
in the computation cost\cite{mohdur}. 
However, when we apply this technique to 
real data, the non-Gaussian nature of the detector 
noise mentioned above 
causes a trouble. As we shall explain later, 
a simple hierarchical search scheme does not work 
in the presence of non-Gaussian noise.  
To solve this difficulty, 
we propose some new computation techniques supplementary to  
the technique of hierarchical search. 
These new techniques depend very much on our choice of 
new mass parameters.  

This paper is organized as follows. 
In Section 2 we explain the definition of our new coordinates 
which parameterize the post-Newtonian templates. 
We also explain that the computation cost 
in the template generation process can be reduced  
by using our new coordinates.  
In Section 3 we discuss a difficulty in 
the hierarchical search, which has not been pointed out  
in literature, and explain a method to 
overcome this difficulty. 
%Only the basic idea of our solution to this problem is explained. 
%The detailed explanation of our scheme for 
%the two step search will be explained elsewhere.  
Section 4 is devoted to summary and discussion. 

Throughout this paper, we use units such that Newton's gravitational
constant and the speed of light are equal to unity. 
The Fourier transform of a function $h(t)$ is denoted by 
$\tilde{h}(f)$, i.e., 
\begin{equation}
\tilde{h}(f):= \int^\infty_{-\infty} dt\, e^{2 \pi i f t} h(t).
\end{equation}

\section{new coordinates for template space}

\subsection{The noise spectrum and templates}

We assume that the time-sequential data 
of the detector output $s(t)$ consists of a signal 
plus noise $n(t)$.
% as 
%\begin{equation}
%s(t) = \tilde\rho h(t) + n(t). 
%\end{equation}
%$Here $\tilde\rho$ is the true amplitude of the signal, and 
We also assume that the wave form of the signal 
is predicted theoretically with sufficiently good accuracy. 
Hence the signal is supposed to be identical to one of 
templates except for the normalization of its amplitude.

To characterize the detector noise, we define one-sided 
power spectrum density $S_n(f)$ by
\begin{equation}
S_n(f)=2\int_{-\infty}^{\infty}\langle n(t) 
 n(t+\tau)\rangle e^{2\pi if\tau} d\tau,
\end{equation}
where $\langle ~~\rangle$ represents the 
operation of taking the statistical average. 
{}For the purpose of the present paper, the overall amplitude of $S_n$ 
is irrelevant. 
We adopt the ``TAMA phase II'' noise spectrum as a model, 
which is given by\cite{TK}
\begin{equation}
S_n(|f|)=\left[{f\over 104\mbox{Hz}}\right]^{-25}
       +\left[{f\over 201\mbox{Hz}}\right]^{-4}
       +1+\left[{f\over 250\mbox{Hz}}\right]^2. 
\end{equation}

%and the other is that for the ``LIGO I noise'', which is given by 
%\begin{equation}
%S_n(|f|)=\left({f\over 175\mbox{Hz}}\right)^{-4}
%       +2\left({f\over 175\mbox{Hz}}\right)^2.
%\end{equation} 

We adopt the templates calculated by using 
the post-Newtonian approximation of general relativity\cite{PN}. 
We use a simplified version of the post-Newtonian templates 
in which the phase evolution 
is calculated to 2.5 post-Newtonian order, but the amplitude 
evolution contains only the lowest Newtonian quadrupole contribution. 
We also use the stationary phase approximation,@
whose validity 
has already been confirmed in Ref.\cite{stp}. 

We denote the parameters distinguishing different templates 
by $M^{\mu}$. 
They consist of the coalescence time 
$t_c$, the total mass $m_{tot} (:=m_1+m_2)$, 
the mass ratio $\eta (:=m_1 m_2/m_{tot}^2)$, 
and spin parameters. 
The templates corresponding to a given set of $M^\mu$ 
are represented in Fourier space by 
two independent templates $\tilde{h}_c$ 
and $\tilde{h}_s$ as 
\begin{equation}
\tilde{h}=\tilde{h}_c \cos\phi_0+\tilde{h}_s \sin\phi_0,
\end{equation}
where $\phi_0$ is the phase of wave, and 
\begin{eqnarray}
\tilde h_c(M^{\mu},f) &= &
i \tilde h_s(M^{\mu},f)=
{\cal N} f^{-7/6} e^{i(\psi_{\theta}(f)+t_c f)},\cr
&&\quad \mbox{for}~~ 0<f\leq f_{max}(M^{\mu}),\cr
 \tilde h_c(M^{\mu},f) & =&\tilde h_s(M^{\mu},f)=0,\cr
&&\quad \mbox{for}~~ f> f_{max}(M^{\mu}).
\end{eqnarray}
Here ${\cal N}$ is a normalization constant, and
\begin{equation}
\psi_{\theta}(f)=\sum_i \theta^i(M^{\mu})\zeta_i(f),
\end{equation}
with
\begin{eqnarray}
 &&\theta^1={3\over 128\eta}(\pi m_{tot})^{-5/3},
\cr
 &&\theta^2={1\over 384\eta}\left({3715\over 84}+55\eta\right)
  (\pi m_{tot})^{-1},
\cr
 &&\theta^3={1\over 128\eta}
  \Bigl((113-86\eta)\chi_s+113\chi_a{m_2-m_1\over m_{tot}}\cr
     &&\hspace*{2cm}-48\pi\Bigr)
  (\pi m_{tot})^{-2/3},
\cr
 &&\theta^4={3\over 128\eta}\Bigl({15293365\over 508032}
   +{27145\over 504}\eta+{3085\over 72}\eta^2\cr
     &&\hspace*{1cm}+\left(30+{275\over4}\eta\right)
   (\chi_s^2-\chi_a^2)\Bigr) (\pi m_{tot})^{-1/3},
\cr
 &&\theta^5={\pi\over 128\eta}
   \left({38645\over 252}+5\eta \right),
\cr
 &&\zeta_1(f)=f^{-5/3},\quad
 \zeta_2(f)=f^{-1},\quad
 \zeta_3(f)=f^{-2/3},\cr
&& \zeta_4(f)=f^{-1/3},\quad
 \zeta_5(f)=\ln f.
\label{eq:temppara}
\end{eqnarray}
We have quoted the expression for the case in which the spin vector of 
each star is aligned or anti-aligned with the axis of the orbital angular 
momentum. 
The spin parameters $\chi_s$ and $\chi_a$ are related 
to the angular momenta of constituent stars $S_{1}$ and $S_2$ by 
\begin{eqnarray}
\chi_s&=&{1\over 2}\left({S_1\over m_1^2}+{S_2\over m_2^2}\right), \\
\chi_a&=&{1\over 2}\left({S_1\over m_1^2}-{S_2\over m_2^2}\right), 
\end{eqnarray}
where a plus (minus) sign is assigned to the angular momentum 
when the spin is aligned (anti-aligned) with the orbital angular momentum. 
In (\ref{eq:temppara}), 
we have neglected the spin effects at 2.5PN order. 
Negative frequency components are given by the reality 
condition of $h(t)$ as 
\begin{equation}
\tilde h(-f)={\tilde h(f)^*},
\end{equation}
where $*$ means the operation of taking the complex conjugate. 

When we consider rather massive 
binaries, $f_{max}$ must be chosen at the frequency  
below which the post-Newtonian
templates are valid. 
On the other hand, when we consider less massive binaries, 
the maximum frequency $f_{max}$ is determined by the 
noise curve alone.  
In this case, we need to 
choose $f_{max}$ such that the loss of the signal-to-noise 
due to the discreteness of 
the time step, $\Delta t_c=1/(2f_{max})$, 
is negligibly small. 

\subsection{Template space in matched filtering and 
new parameters}

Here, we define the inner product between
two real functions $a(t)$ and $b(t)$ as \cite{CF}
\begin{eqnarray}
(a,b)&=&\int^\infty_{-\infty}df {\tilde{a}(f) \tilde{b}^*(f)
\over S_n(|f|)}. 
%\nonumber\\
%&=&
%\int^\infty_0 {{\tilde{a}(f) \tilde{b}^*(f)+
%\tilde{a}^*(f) \tilde{b}(f)} \over S_n(|f|)}.
\label{innerproduct}
\end{eqnarray}

In the matched filtering, 
we define the filtered signal-to-noise ratio after 
maximization over $\phi_0$ as 
\begin{equation}
\rho=\sqrt{(s,h_c)^2+(s,h_s)^2}. 
\end{equation}
We choose the normalization constant ${\cal N}$ to satisfy
$(h_c,h_c)=$ $(h_s,h_s)=1$.  
%\trg{Then, in the presence of \er{Gaussian} noise alone ($\tilde\rho=0$), 
%the root mean square of $\rho$ becomes $\sqrt{2}$.}{}

Since the best fit value for parameters are not known in advance,
we must filter the data through many templates 
at different points in the parameter space. 
In order to determine representative points in the parameter space, 
we have to know how much the value of $\rho$ is reduced 
by using a template with different mass parameters 
from the best ones. 
Here, we adopt geometrical description 
of the template space\cite{owen}, and investigate 
which coordinates we should choose to simplify 
the strategy for determining representative points
in the parameter space.  

In the following, we assume that 
the maximum frequency $f_{max}$ is determined 
by the noise curve alone independently 
of the template parameters. 
The effect of parameter dependence of $f_{max}$ is 
discussed at the end of this section. 
Although $\theta^i$ introduced above are functions of $M^\mu$, 
we regard them as independent variables for a while. 
Hence, we parameterize templates like $\tilde h_c(\theta')$, 
where we defined $\theta'$ as the set of parameters $(t_c,\theta)$ 
setting $\theta'_0=t_c$ and $\theta'_i=\theta_i$ for $i=1,\cdots,5$.

The correlation between two nearby 
templates with different $\theta'$ 
is evaluated as 
\begin{eqnarray}
 &&(\tilde h_c(\theta'+\Delta\theta'),\tilde h_c(\theta'))\cr
 &&\quad=2{\cal N}^2 \int_{0}^{f_{max}} 
  {df\over S_n(f)} f^{-7/3} \cos(\Delta \psi_{\theta}(f)+\Delta t_c f), 
\end{eqnarray}
where $\Delta \psi_{\theta}(f)= 
\psi_{\theta+\Delta\theta}(f)-\psi_{\theta}(f)$.
In the same manner, we have
\begin{eqnarray}
 &&(\tilde h_c(\theta'+\Delta\theta'),\tilde h_s(\theta'))\cr
 &&\quad =2{\cal N}^2 \int_{0}^{f_{max}} 
  {df\over S_n(f)} f^{-7/3} \sin(\Delta \psi_{\theta}(f)+\Delta t_c f).
\end{eqnarray}
Here it should be emphasized that
these correlations depend on $\theta'$ 
only through $\Delta\theta'$. 

Let us define new functions ${\cal G}'(\Delta\theta')$ 
and ${\cal G}(\Delta\theta)$ by 
\begin{eqnarray}
{\cal G}'(\Delta\theta')&:=&
[(\tilde h_c(\theta'+\Delta\theta'),\tilde h_c(\theta'))^2\cr
&&\quad +(\tilde h_c(\theta'+\Delta\theta'),\tilde h_s(\theta'))^2]^{1/2}, 
\end{eqnarray}
and ${\cal G}(\Delta\theta):=\max_{\Delta t_c}{\cal G}'(\Delta\theta')$, 
respectively. 
This function ${\cal G}$ is known as the {\it match}. 
We expand ${\cal G}'(\Delta\theta')$ with respect to 
$\Delta\theta'$ as 
\begin{eqnarray}
{\cal G}'(\Delta\theta')&=&
\Biggl[1-{1\over 2}{\cal N}^2
[[f^{-7/3}(\Delta \psi_{\theta}(f)+\Delta t_c f)^2]]\cr
&&\quad+{1\over 2}{\cal N}^4
[[f^{-7/3}(\Delta \psi_{\theta}(f)+\Delta t_c f)]]^2
+\cdots \Biggr], \hspace{-1cm}\cr &&
\end{eqnarray}
where we introduced a notation 
\begin{equation}
 [[g(f)]]:=\int_{0}^{f_{max}}{df\over S_n(f)} (g(f)+g^*(f)).
\end{equation}
We define a matrix $G'_{\mu\nu}$ by 
\begin{equation}
{\cal G}'(\Delta\theta')=
1-G'_{\mu\nu}\Delta\theta'{}^\mu
\Delta\theta'{}^\nu+\cdots.
\end{equation}
By definition, 
this matrix $G'_{\mu\nu}$ is a constant 
matrix independent of $\theta'$, and 
it is determined once the noise spectrum is specified. 
In order to take maximum of ${\cal G}'$ with respect 
$\Delta t_c$, we project $G'_{\mu\nu}$ on to the 
space orthogonal to $t_c$ as 
\begin{equation}
G_{ij}=G'_{ij}-{G'_{i0} G'_{j0}\over G'_{00}}.
\end{equation} 
The matrix $G_{ij}$ can be considered as a five dimensional metric 
analogous to the two dimensional one introduced in Ref.~\cite{owen}.

Next, we orthonormalize $G_{ij}$ as 
\begin{equation}
G_{ij}=\sum_{\alpha=1}^5\lambda_{\alpha} P^{\alpha}{}_i P^{\alpha}{}_j, 
\end{equation}
where $\lambda_{\alpha}$ are the eigenvalues and $P^{\alpha}{}_i$
is an orthogonal matrix composed of the eigen vectors of $G_{ij}$. 
Rotating the axis further by using another orthogonal 
matrix $Q^A{}_{\alpha}$, we define new coordinates 
of the five dimensional template space by 
\begin{equation}
x^A :=\sum_{\alpha=1}^{5}\sum_{i=1}^{5} 
Q^A{}_{\alpha}\lambda_\alpha^{1/2} P^{\alpha}{}_{i}\theta^i, 
\quad (A=1,\cdots,5). 
\label{defx}
\end{equation}

Let us denote our five dimensional template space 
as $\Gamma$. 
In this paper, we assume that we can neglect the effect of spins 
of binary stars. 
Therefore, the actual template space to be searched 
becomes a two dimensional hypersurface $S$ in $\Gamma$. 
Since $\theta^i$ are functions of $m_{tot}$ and $\eta$, 
we find that Eq.(\ref{defx}) defines  
a map from the actual template space parameterized by $(m_{tot},\eta)$ 
to $\Gamma$. 
Then, this map naturally specifies $S$.  

One of the most important points that we wish to emphasize 
in this paper is 
that the geometry of this two dimensional surface $S$ 
becomes almost flat. 
Because of this fact, 
we can choose $Q^A{}_{\alpha}$ so that 
the $x^1$ and $x^2$ axes 
lie approximately on $S$. 
Taking into account 
the extension of the area to be searched on $S$, 
we choose $Q^A{}_{\alpha}$ 
by solving the following set of equations,  
\begin{eqnarray}
 && {\bf x}(m_{min},m_{min})-{\bf x}(m_{max},m_{max})
\cr&&\quad\quad\quad\quad\quad\quad\quad\quad\quad\quad
  =(\alpha_{11},0,0,0,0), 
\cr
 && {\bf x}(m_{min},m_{min})-{\bf x}(m_{min},m_{max})
\cr&&\quad\quad\quad\quad\quad\quad\quad\quad\quad\quad
  =(\alpha_{21},\alpha_{22},0,0,0),
\end{eqnarray}
where $m_{min}$ and $m_{max}$ are 
the minimum and the maximum mass of templates, respectively. 
The directions of the other axes are not important here. 
(Hence, we do not specify how to choose them explicitly. )

By solving $X^1=x^1(m_{tot},\eta)$ and $X^2=x^2(m_{tot},\eta)$ for 
$m_{tot}$ and $\eta$, we obtain inverse functions $m_{tot}(X^1,X^2)$ and 
$\eta(X^1,X^2)$, and we can use $X^1$ and $X^2$ as parameters 
for the template space 
instead of $m_{tot}$ and $\eta$. 
Furthermore, we can define a 
map from $(X^1,X^2)$ to $\Gamma$ by 
$x^A=x^A(m_{tot}(X^1,X^2),\eta(X^1,X^2))$ for A=3,4,5. 

In the following, we verify that 
the two dimensional hypersurface $S$ is almost flat. 
First we check that the parameters $x^3$, $x^4$, and $x^5$ 
are approximately zero on any points on $S$. 
To show this, we plot the value of 
$\delta x:=\sqrt{(x^{3})^2+(x^4)^2+(x^5)^2}$ as a 
function of $m_{1}$ and $m_2$ in Fig.1. 
Here we set $m_{min}=0.2 M_{\odot}$ and $m_{max}=3 M_{\odot}$. 
We find that $\delta x$ are very small. 
This indicates that the surface $S$ is almost flat and  
$(X^1,X^2)$ can be regarded as Cartesian coordinates on $S$ 
in a good approximation. 

The metric associated with the new coordinates 
is defined by 
\begin{eqnarray}
 g_{IJ}&:=&
  \sum_{i,j=1}^{5} G_{ij}{\partial \theta^i\over \partial X^{I}}
           {\partial \theta^j\over \partial X^{J}}\cr
&=& \sum_{A,B=1}^{5}
\delta_{AB}\sum_{\mu=1}^{5}\left({\partial x^A\over \partial M^{\mu}}
  {\partial M^{\mu}\over \partial X^I}\right)
   \sum_{\nu=1}^{5} \left({\partial x^B\over \partial M^{\nu}}
  {\partial M^{\nu}\over \partial X^J}\right), 
\hspace{-1cm}\cr
&&
(I,J=1,2). 
\end{eqnarray}
To show the constancy of this metric, we plot the residuals 
$g_{IJ}-\delta_{IJ}$ in Fig.2(a)-(c).

\begin{figure*}
\label{Fig.1}
\centerline{\epsfysize=5cm 
\epsfbox{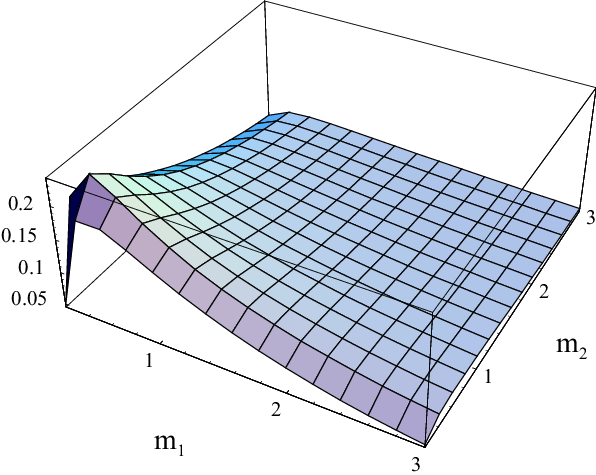}} \caption{
The value of $\delta x$ 
%${\sqrt{(x^3)^2+(x^4)^2+(x^5)^2}}$ 
as a function of $m_1$ and $m_2$ in solar mass unit.}
\end{figure*}

\begin{figure*}
\epsfysize= 5cm 
\centerline{\epsfbox{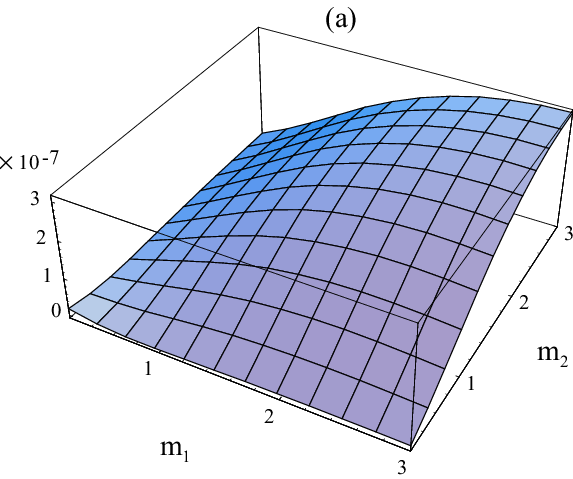}} 
\epsfysize= 5cm 
\centerline{\epsfbox{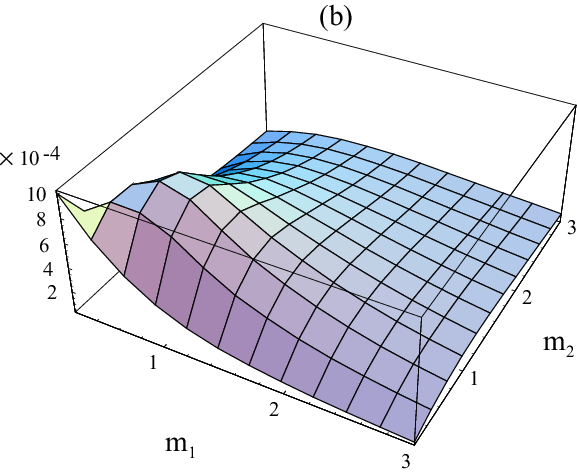}} 
\epsfysize= 5cm 
\centerline{\epsfbox{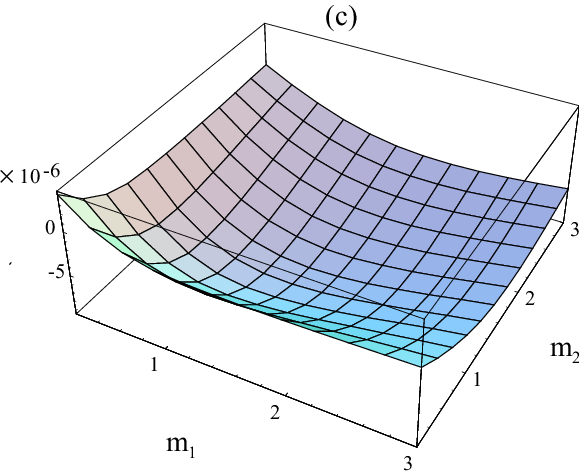}} 
\caption{The values of (a)$g_{11}-1$, (b)$g_{22}-1$, 
(c)$g_{12}$, as functions of $m_1$ and $m_2$
in solar mass unit. 
}
\label{Fig.2}
\end{figure*}
\newpage

These facts suggest the usefulness 
of the new coordinates. 
First of all, the flatness of the metric allows us to use a uniform square 
grid to generate the template bank. 
Besides, there are several advantages. 
As long as we consider a small area in $\Gamma$,  
$x^3, x^4$ and $x^5$ can be treated as constants. 
We can make use of this fact to develop 
an efficient algorithm to generate templates in frequency domain 
as we shall see in the succeeding subsection.

\subsection{An efficient algorithm to generate templates}

We can express $\theta^i$ as 
linear functions of $x^A$ by solving 
Eq.(\ref{defx}) inversely. Thus, the phase function 
$\psi_\theta(f)$ is also a linear function of $x^A$. 
Since the effect of variation of $(x^3,x^4,x^5)$ within 
a small area is negligibly small, 
the difference of the phase function 
\begin{equation}
\Delta\psi(f)=
\psi_{(X^1+\Delta X^1,X^2+\Delta X^2)}(f)-\psi_{(X^1,X^2)}(f)
\end{equation} 
is almost independent of $(X^1,X^2)$. 
Therefore, we can prepare the phase difference 
$e^{i\Delta\psi(f)}$ for various values of $(\Delta X^1,\Delta X^2)$
in advance.  
Then we can calculate the template 
$\tilde h_{(X^1+\Delta X^1,X^2+\Delta X^2)}(f)$ just by 
multiplying the corresponding phase difference by 
the template at $(X^1,X^2)$ as 
\begin{equation}
\tilde h_{(X^1+\Delta X^1,X^2+\Delta X^2)}(f)=
\tilde h_{(X^1,X^2)}(f)\, e^{i\Delta\psi(f)}. 
\end{equation}
Hence, once we calculate one template, 
we do not have to call subroutines of the 
sinusoidal functions to generate neighboring templates of it. 
Since the computation of sinusoidal
functions is slow in many compiler, this algorithm 
significantly reduces the computation cost to generate templates.

Before closing this secton, 
we remark on the choice of the maximum frequency
$f_{max}$. 
So far, we have been neglecting the fact 
that, in general, the maximum frequency 
$f_{max}$ depends on the template parameters. 
When we consider binaries with large mass, 
the frequency at the last stable circular orbit 
becomes lower than the maximum frequency 
which is determined by the shape of the noise power 
spectrum. Since our post-Newtonian templates 
are no longer valid beyond the last stable orbit, 
the maximum frequency should be chosen below the 
corresponding frequency. 
Even in that case, we think that our new coordinates are still 
useful by the following reason. 
The {\it match} determined with larger 
$f_{max}$ is likely to underestimate the correct value. 
Hence, the grid spacing determined by 
using our new coordinates tends to be closer 
than that determined by more accurate estimation. 
Therefore, to adopt constant $f_{max}$ in determining 
the template bank would be safe in the sense that it is less likely 
to miss detectable events. 
Although the number of templates increases with the choice 
of constant $f_{max}$, 
such effect is very small. This is because 
the number of templates with relatively large mass 
is not very large. 
Recall that the number of template are dominated by 
templates with small mass $m\sim m_{min}$\cite{owen2}. 

\section{new fast algorithm for hierarchical search}
If we try to search gravitational waves from 
coalescing binaries with mass $M\geq 0.2 M_{\odot}$, 
we have to calculate the correlations $(\tilde s(f),\tilde h(f))$ 
for several$\times 10000$ templates with 
different mass parameters\cite{owen2}
\footnote{
If we take into account the effect of spins of binary stars, this 
number will increase about 3 times or more. We will discuss 
this issue in a separate paper\cite{paper2}.
}. 
The computation cost 
to evaluate such a large number of correlations is 
very expensive. 
One promising idea to reduce the computation cost 
is the technique of hierarchical search. 
The basic idea of hierarchical search is as follows. 
At the first step, we examine the correlations with 
a smaller number of templates located more sparsely. 
In order not to lose the candidates of events, 
we set a sufficiently low threshold of the filtered signal-to-noise $\rho$ 
at the first step. 
If a set of parameters $(X^1,X^2,t_c)$ is 
selected as a candidate, 
we examine the correlations 
between the data and the neighboring templates 
by using a finer mesh. 

A simple scheme for two step search has been already 
discussed in Ref.\cite{mohdur}. 
However, there seems to be a problem 
in realizing the basic idea mentioned above.  
It has been pointed out that the 
distribution of the amplitude of the detector noise 
will not follow the simple stationary Gaussian statistics\cite{allen}.
The non-stationary and non-Gaussian nature of the detector noise 
will produce a lot of events with 
large value of $\rho$. 
Namely, there seems to exist a non-Gaussian tail part in the 
distribution of $\rho$. 
In order to identify real events, 
we need to 
keep the expected number of fake events small 
by choosing the threshold of $\rho$ as being sufficiently large. 
Hence, the existence of 
the non-Gaussian tail requires 
a larger value of threshold of $\rho$. 
This leads to a significant loss of detector sensitivity. 
To avoid such a loss, 
it was proposed to use a $\chi^2$-test as a supplementary 
criterion. 

Here, $\chi^2$ is defined as follows. 
First we divide each template into mutual independent $n$ pieces, 
\begin{equation}
 \tilde h_{(c,s)}(f)=\tilde h^{(1)}_{(c,s)}(f)
+\tilde h^{(2)}_{(c,s)}(f)+
             \cdots +\tilde h^{(n)}_{(c,s)}(f), 
\end{equation}
and we calculate 
\begin{equation}
 z^{(i)}_{(c,s)}=(\tilde s,\tilde h^{(i)}_{(c,s)}), 
\quad 
 \bar z^{(i)}_{(c,s)}=\sigma_{(i)}^2\times (\tilde s,\tilde h_{(c,s)}), 
\end{equation}
with
\begin{equation}
\sigma_{(i)}^2=(\tilde h^{(i)}_{(c)},\tilde h^{(i)}_{(c)})
=(\tilde h^{(i)}_{(s)},\tilde h^{(i)}_{(s)}).
\end{equation}
Then $\chi^2$ is defined by 
\begin{equation}
\chi^2=\sum_{i=1}^n \left[
{\left(z^{(i)}_{(c)}- {\bar z^{(i)}_{(c)}}\right)^2 + 
\left(z^{(i)}_{(s)}- {\bar z^{(i)}_{(s)}}\right)^2
\over \sigma_{(i)}^2}\right]. 
\end{equation}
This quantity must satisfy the $\chi^2$-statistics with 
$2n-2$ degrees of freedom and must be independent 
of $\rho=\sqrt{z_{(c)}^2+z_{(s)}^2}$, 
as long as the detector noise is Gaussian. 
However, as reported in Ref.~\cite{allen}, 
events with large $\chi^2$ in reality occur 
more often than in the case of Gaussian noise. 
There is a strong tendency that events with large $\chi^2$ have 
a large value of $\rho$ on average. 
Thus, by changing the threshold of $\rho$ depending on the value 
of $\chi^2$, we can reduce the number of fake events without 
any significant loss of detector sensitivity. 
Hence, it will be necessary to implement the $\chi^2$-test 
even in the simple 
one step search case.
%\footnote{
%We have developed a C-code to implement 
%the $\chi^2$-test in the two step search 
%for the data analysis of TAMA300. 
%The detailed strategy specific to this code will 
%be reported at some other place.} 
However, if we try to evaluate $\chi^2$, 
the computation cost becomes more 
expensive\footnote{
If we try to evaluate $\chi^2$ naively, 
the computation cost necessary for the second step search 
simply becomes $n$ times larger. 
Since $n$ will be chosen as being $O(10)$, 
the increase of the computation cost is unacceptable. 
If one calculates the values of $\chi^2$ only for a few varieties of 
coalescence time at which a large value of $\rho$ is 
achieved, the computation 
cost for $\chi^2$ might be kept small. 
In this case, we can use the direct summation 
instead of FFT to calculate the values of $\chi^2$. 
But, the question is for how many 
varieties of coalescence time we must calculate 
the values of $\chi^2$ not to lose real events.  
This is not a simple question. 
If this number is sufficiently small,  
this naive strategy will work in the case of one step search. 
}.
Thus, it is strongly desired to implement 
an efficient algorithm  
to reduce the computation cost.  

Now, we discuss a method of two step hierarchical search 
taking into account the presence of non-Gaussian noise. 
At the first step search, 
a large number of candidates 
for the second step search with large $\rho$ value 
will appear due to the non-Gaussianity of noise. 
As is mentioned above, 
in order not to lose the detector sensitivity, 
it is desired to introduce the $\chi^2$-test, and 
to keep the threshold of $\rho$ small. 
Furthermore, by introducing the $\chi^2$-test at the first step,  
we can reduce the number of candidates for the second step search. 
Thus, the $\chi^2$-test is also effective to reduce the computation 
cost for the second step. 
However, the $\chi^2$-test at the first step
increases the computation cost for the first step. 
This increase in the computation cost can be very large 
in the presence of non-Gaussian noise  
because the number of fake events which exceed the 
threshold of $\rho$ at the first step 
is much larger than that expected
in the case of Gaussian noise.  
Then, we must compute a lot of $\chi^2$ 
values at the first step. 
When we take into account these effects, 
the advantage of the two step search, which 
is estimated to be about factor $30$ in comparison with 
the simple one step search in the case of Gaussian 
noise\cite{mohdur}, will be significantly reduced 
or will be totally lost. 
Hence, in order to make use of the potential advantage 
of the two step search, 
we need other ideas to reduce the computation cost further. 
Here we present two new ideas of this kind.

\vbox{
\vspace*{5mm}
\hspace*{3mm}(a)
\vspace{-8mm}
\begin{center}
$\to X^1$
\vspace*{2mm}

$\begin{array}{c}
X^2\\
\uparrow
\end{array}
$\hspace{-4mm}
\begin{tabular}{c|ccccc}
~~~~~~~~~~ & ~~0.0~~ & ~~0.5~~ & ~~1.0~~ & ~~1.5~~ 
 & ~~2.0~~ \\ 
\hline
 2.0 & 0.774 & 0.760 & 0.757 & 0.746 & 0.717\\
 1.5 & 0.828 & 0.800 & 0.777 & 0.765 & 0.733\\
 1.0 & 0.889 & 0.857 & 0.804 & 0.771 & 0.733\\
 0.5 & 0.947 & 0.918 & 0.821 & 0.742 & 0.682\\
 0.0 & 1.000 & 0.914 & 0.765 & 0.644 & 0.592\\ 
-0.5 & 0.947 & 0.886 & 0.774 & 0.653 & 0.553\\
-1.0 & 0.889 & 0.865 & 0.791 & 0.692 & 0.589\\
-1.5 & 0.828 & 0.831 & 0.794 & 0.724 & 0.637\\
-2.0 & 0.774 & 0.787 & 0.777 & 0.737 & 0.674\\
\end{tabular}
\end{center}}

\vbox{
\hspace*{3mm}(b)
\vspace{-4mm}
\begin{center}
%$\to X^1$
%\vspace*{2mm}

$\begin{array}{c}
X^2\\
\uparrow
\end{array}
$\hspace{-4mm}
\begin{tabular}{c|ccccc}
~~~~~~~~~~ & ~~0.0~~ & ~~0.5~~ & ~~1.0~~ & ~~1.5~~ 
 & ~~2.0~~ \\ 
\hline
2.0 & 0.781 & 0.759 & 0.770 & 0.747 & 0.735\\
1.5 & 0.838 & 0.817 & 0.797 & 0.766 & 0.739\\ 
1.0 & 0.904 & 0.873 & 0.812 & 0.764 & 0.710\\ 
0.5 & 0.961 & 0.907 & 0.809 & 0.728 & 0.673\\ 
0.0 & 0.984 & 0.911 & 0.791 & 0.663 & 0.608\\
-0.5& 0.961 & 0.893 & 0.777 & 0.666 & 0.565\\ 
-1.0& 0.904 & 0.869 & 0.803 & 0.699 & 0.598\\
-1.5& 0.838 & 0.841 & 0.809 & 0.732 & 0.646\\
-2.0& 0.781 & 0.803 & 0.787 & 0.750 & 0.681\\
\end{tabular}
\end{center}
}
\noindent
\noindent
{\small TABLE 1. 
Tables of maximum correlations for various choices 
of $\Delta {\bf X}$ (a) with 5000Hz sampling  
and (b) with 1250Hz sampling.}
\vspace{5mm}

The first one is very simple. 
At the first step search, we can reduce the sampling rate of 
the data. The low sampling rate results in  
the reduction in the filtered signal-to-noise mainly due to 
the mismatch in $t_c$. 
However such reduction can be compensated 
by a very small change of the threshold of $\rho$. 
In the case of the ``TAMA noise curve'',
we can allow the sampling rate as low as about 1000Hz. 
The values of {\it match} between two templates with various 
$\Delta {\bf X}$ are shown in Table.1(a), 
where we adopted 5000Hz as the sampling rate. 
The same quantities for 1250Hz sampling are 
shown in Table.1(b). 
Here, one of the templates is considered as a normalized signal 
without noise, and the other as a search template. 
The signal is normalized to satisfy $(h,h)=1$ with $f_{max}=2500Hz$ 
in both cases of Table.1(a) and Table.1(b).  
On the other hand, the search template is normalized 
with $f_{max}=2500Hz$ in the case of Table.1(a)  
and with $f_{max}=625Hz$ in the case of Table.1(b). 
We find that the difference of the values of {\it match} between 
these two cases are very small 
especially for large $|\Delta {\bf X}|$. 
Therefore, detection probability for a fixed threshold 
of $\rho$ is not significantly 
lost even if we adopt a rather small sampling rate 
at the first step search, 
at which we adopt a relatively large spacing for the template bank. 
The reduction in the sampling rate directly reduces the 
computation cost. 
The usual FFT routine requires effective  
floating point operations proportional to $\sim N\ln N$ 
to compute the Fourier transform of the data with length $N$. 
Furthermore, for most of FFT routines and computer environments, 
the effective FLOPS value for FFT is larger for smaller $N$. 
Thus, the reduction factor for the computation cost   
due to adopting a smaller FFT length is much larger than 
one expects naively.

The second idea is more important. 
What we need to evaluate is the correlation 
\begin{eqnarray}
 Z=(\tilde h,\tilde s) 
  & =&\int_{-f_{max}}^{f_{max}} df
  {\tilde h_{\bf X}(f) \tilde s^{*}(f)\over S_n(f)} \cr
  & \approx&{\cal N}\Delta f \sum_{j=1}^N 
  \left[{\tilde s^{*}(f_j) e^{i\psi_{\bf X}(f_j)} 
          \over f_j^{7/6} S_n(f_j)}\right]
    e^{2\pi i f_j t_c}. 
\end{eqnarray}
where we take it into account that in reality 
we deal with a discrete time sequence of data 
with length $N$. 
$\Delta f$ is given by the sampling rate divided by $N$, 
and $f_j:=\Delta f(j-N/2)$. 
The correlations for various values of $t_c$ are 
calculated simultaneously 
by taking the Fourier transform of the array defined by 
the quantity inside 
the square brackets in the last line of the above equation. 
This scheme is efficient enough when we do not have any guess about 
$t_c$. However, when we perform the second step search 
we have a good estimate of $t_c$ 
at which the maximum correlation is expected to be achieved. 
We denote it by $\hat t_c$. 
In this case, we need to evaluate $Z$ only for $t_c$ 
which are close to $\hat t_c$. 
Also for templates, we have a good guess for the mass parameters, $\hat{\bf X}$.  
Thus, we need to calculate the correlation 
$Z$ only for a cluster of the templates neighboring to $\hat{\bf X}$.

Once $\hat t_c$ and $\hat{\bf X}$ are specified, 
we can rewrite the above expression 
in a very suggestive form as 
\begin{equation}
 Z \approx {\cal N}\Delta f \sum_{k=1}^{N/m-1} 
 \sum_{j=m(k-1/2)+1}^{m(k+1/2)} A_j \times B_j,
\label{AB}
\end{equation}
with
\begin{eqnarray}
  A_j & =& {\tilde s^{*}(f_j) e^{i\psi_{\hat{\bf X}}(f_j)
         +2\pi i f_j \hat t_c} 
          \over f_j^{7/6} S_n(f_j)},\cr
  B_j & =& e^{i\Delta\psi(f_j)+2\pi i f_j \Delta t_c}.
\end{eqnarray}
Here $\Delta\psi(f_j):=\psi_{\bf X}(f_j)-\psi_{\hat{\bf X}}(f_j)$   
and $\Delta t_c:=t_c-\hat t_c$. 
We have also introduced $m$ as a certain integer which divides $N$.
As long as both $|\bf X-\hat{\bf X}|$ and $|t_c-\hat t_c|$
are sufficiently small, the factor $B$ is a slowly changing 
function of frequency. Hence, unless $m$ is not large, 
$B$ can be moved outside the second summation 
in Eq.(\ref{AB}). 
Then, introducing 
\begin{equation}
 A'_k=\sum_{j=m(k-1/2)+1}^{m(k+1/2)} A_j, 
\end{equation}
we obtain 
\begin{eqnarray}
 Z &\approx& {\cal N}\Delta f \sum_{k=1}^{N/m-1} A'_k \times B_{mk}\cr 
     & = & {\cal N}\Delta f \sum_{k=1}^{N/m-1} 
        \left[A'_k  e^{i\Delta\psi(f_{mk})}\right] 
        \times e^{2\pi i f'_k\Delta t_c}, 
\label{zfinal}
\end{eqnarray}
where $f'_k:= m\Delta f\left(k-{N\over 2m}\right)$. 
The expression in the last line can be 
evaluated by applying the FFT routine to 
the array defined by the quantity 
inside the square brackets. 
The correlation between two templates 
for various values of $\Delta {\bf X}$ and $\Delta t_c$ 
are calculated by using this method. 
The results are shown in Table.2 for $N/m=1024, 2048$ and $4096$. 
We find that $N/m$ can be taken as small as $2048$ without 
significant loss in accuracy.

\begin{center}

\vspace*{5mm}
(a)
\vspace*{2mm}

\begin{tabular}{|c|c|c|c|c|}
\hline
$~~\Delta t_e$(sec)~~ & $~~m=4096~~$ & $~~m=2048~~$ 
& $~~m=1028~~$
\\
\hline
0.0000 & 1.000 & 1.000 & 1.000 
\\
\hline
0.0128 & 1.000 & 0.999 & 0.994 
\\
\hline
0.0248 & 0.999 & 0.994 & 0.975 
\\
\hline
\end{tabular}

\vspace*{5mm}
(b)
\vspace*{2mm}

\begin{tabular}{|c|c|c|c|c|}
\hline
$~~\Delta t_e$(sec)~~ & $~~m=4096~~$ & $~~m=2048~~$ 
& $~~m=1028~~$
\\
\hline
0.0000 & 0.765 & 0.765 & 0.765 
\\
\hline
0.0128 & 0.764 & 0.763 & 0.760 
\\
\hline
0.0248 & 0.763 & 0.760 & 0.746 
\\
\hline
\end{tabular}

\vspace*{5mm}
(c)
\vspace*{2mm}

\begin{tabular}{|c|c|c|c|c|}
\hline
$~~\Delta t_e$(sec)~~ & $~~m=4096~~$ & $~~m=2048~~$ 
& $~~m=1028~~$
\\
\hline
0.0000 & 0.774 & 0.774 & 0.774 
\\
\hline
0.0128 & 0.774 & 0.773 & 0.769 
\\
\hline
0.0248 & 0.773 & 0.769 & 0.755 
\\
\hline
\end{tabular}

\end{center}

\noindent
{\small TABLE 2. 
Tables of maximum correlations for various choices of $m$ 
and $\Delta t_e=\hat t_e-t_e^{(c)}$ with 
(a) $\Delta {\bf X}=(0,0)$, (b)$\Delta {\bf X}=(1,0)$
and (c)$\Delta {\bf X}=(0,2)$.}

\vspace{5mm}

Furthermore, as an advantage of our new coordinates, 
the factor $e^{i\Delta\psi(f_{mk})}$ can be well approximated 
by the one obtained by setting $\Delta x^3= \Delta x^4=\Delta x^5=0$. 
It means that this factor is almost independent of the values of 
$(X^1,X^2)$. 
Thus, we have to calculate this factor only once 
at the beginning of the second step search. 
Since the array $A'_k$ is independent of $\Delta{\bf X}$,
the quantity inside the square brackets in Eq.~(\ref{zfinal}) 
for various values of $\Delta{\bf X}$ is simply given by 
$A'_k$ times the pre-calculated factor 
$e^{i\Delta\psi(f_{mk})}$.
This fact manifestly leads to an 
additional reduction in the computation cost. 

The same technique can be used to evaluate $\chi^2$, 
i.e., $Z^{(i)}$,  just by replacing the array $A'_k$ with 
the same quantities multiplied by an appropriate window function. 

\section{conclusion}

We discussed a method of 
analyzing data from interferometric gravitational wave detectors  
to detect gravitational waves from inspiraling compact binaries 
based on the technique of matched filtering.  
We described a brief sketch of 
several new techniques which would be useful in 
hierarchical search of gravitational waves.

First, we proposed new parameters 
which label templates of gravitational waves 
from inspiraling binaries. 
These new parameters are chosen so that the 
metric on the template space 
becomes almost constant. 
We found that the template space 
can be well approximated by two dimensional flat Euclidean metric. 
Thus, by using these parameters as coordinates for the
template space, 
the problem of the template placement becomes very simple. 
Say, we can use a simple uniform square grid 
to specify the grid points for the bank of templates. 
Furthermore, we found that, by using new parameters,
we can introduce an efficient 
method to generate templates in frequency domain.  
The reduction in the computation cost is achieved  
by using the property of our new coordinates 
that one template can be translated 
into another with different mass parameters 
by just multiplying pre-calculated 
coefficients to the original template. 
Therefore, we can generate a set of templates from one 
template avoiding calculation of 
the sinusoidal functions. 

Next, we discussed a method of two step hierarchical search. 
Due to the non-stationary and non-Gaussian nature of 
the detector noise,  
we will have to introduce a $\chi^2$-test 
when we analyze real data. 
When we take into account this fact, 
it becomes very difficult to obtain large reduction in the 
computation cost by applying 
naive two-step hierarchical search strategy. 
To solve this difficulty, 
we proposed two new techniques 
to reduce the computation cost 
in the two step search. 
One is to use a lower sampling rate for the first 
step search. By using this technique, 
we can reduce the length of FFT by factor 2 or 4 
keeping the loss of correlation within an acceptable level. 
The second technique, which is more important, 
makes use of the fact that a good guess for 
the coalescence time and the mass parameters 
has been obtained as a result of the first step search 
at the time when we perform the second step search. 
We found that the length of FFT for the second step search 
can be reduced down to about 2048. 

Based on the new techniques discussed in this paper, 
we have developed a 
hierarchical search code to analyze data from 
the TAMA300 detector. 
The details of this code and the result of the analysis of 
the first TAMA300 data will be 
presented elsewhere. 

\acknowledgements
{T.T. thanks B. Allen and A. Wiseman for their 
useful suggestions and encouragements 
given during his stay in Milwaukee at the beginning of this study. 
We thank P. Brady, N. Kanda and M. Sasaki for discussion. 
We also thank K. Nakahira 
for her technical advice in the development of our 
computer code.  
This work is supported in part by 
Monbusho Grant-in-Aid 11740150 and by 
Grant-in-Aid for Creative Basic Research 09NP0801.
Some of the numerical calculation were done by using 
the gravitational wave data analysis library GRASP\cite{grasp}. 
}


\begin{references}
\bibitem{TK} This is an analytic fit to the TAMA phase II 
sensitivity provided to us by D.Tatsumi and N.Kanda (private 
communication).
\end{references}
\end{document}